\documentclass[prb,aps,longbibliography,floats,superscriptaddress,twocolumn]{revtex4-2}

\usepackage[utf8]{inputenc}
\usepackage{amsmath,amssymb,amsthm,amsbsy,bm}
\usepackage{graphicx,epsfig,epstopdf,subfigure}
\usepackage{dsfont,verbatim,varwidth,dcolumn}
\usepackage[usenames,dvipsnames]{xcolor}
\usepackage{color,soul,subfiles}
\usepackage[colorlinks]{hyperref}

\def\ua{{\uparrow}}
\def\da{{\downarrow}}
\newcommand{\gt}{g_{\text{2-ter}}}

\makeatother

\begin{document}

\title{Edge Reconstruction and Emergent Neutral Modes in Integer and Fractional Quantum Hall Phases}

\author{Udit Khanna}
\affiliation{Department of Physics, Bar-Ilan University, Ramat-Gan 52900, Israel}

\author{Moshe Goldstein}
\affiliation{Raymond and Beverly Sackler School of Physics and Astronomy, Tel-Aviv University, Tel Aviv, 6997801, Israel}

\author{Yuval Gefen}
\affiliation{Department of Condensed Matter Physics, Weizmann Institute of Science, Rehovot 76100, Israel}

\begin{abstract}
This paper comprises a review of our recent works on fractional chiral modes that emerge due to edge reconstruction 
in integer and fractional quantum Hall (QH) phases. The new part added is an analysis of edge reconstruction of 
the $\nu = 2/5$ phase. QH states are topological phases of matter featuring chiral gapless modes at the edge. 
These edge modes may propagate downstream or upstream, and may support either charge or charge-neutral 
excitations. From topological considerations, particle-like QH states are expected to support only downstream charge modes. 
However the interplay between the electronic repulsion and the boundary confining potential may drive 
certain quantum phase transitions (called reconstructions) at the edge, which are associated to the nucleation of 
additional pairs of counter-propagating modes. Employing variational methods, here we study edge reconstruction in 
the prototypical particle-like phases at $\nu = 1, 1/3$ and $2/5$ as a function of the slope of the confining potential. 
Our analysis shows that subsequent renormalization of the edge modes, driven by disorder-induced tunnelling and 
intermode interactions, may lead to the emergence of upstream neutral modes. These predictions may be tested in 
suitably designed transport experiments. Our results are also consistent with previous observations of upstream neutral modes
in these QH phases, and could explain the absence of anyonic interference in electronic Mach-Zehnder setups. 
\end{abstract}

\maketitle

{\it Mark Azbel was a physicist in his heart and soul. Beside his major contributions to the quantum theory of electrons 
in metals, he left his footprints everywhere through talks, arguments and discussions on a broad spectrum of issues (including 
non-physics themes). He would take an issue that appears all but benign, and expose the intricacy, deep significance, and a
scale of insight needed to really penetrate the problem at hand. Among his many notable works was the prediction of the 
anomalous skin effect in metals~\cite{Azbel2,Azbel0}. It is only natural to devote our manuscript in this commemorative 
volume to a study of another ``skin'' effect related to the boundary of a two-dimensional electron gas in the presence 
of a strong perpendicular magnetic field. This concerns the gapless chiral edge modes which emerge at the boundary of 
quantum Hall phases.}

\section{Introduction}

Quantum Hall (QH) phases are two-dimensional topological phases featuring gapless, chiral modes 
localized at the boundary of the sample~\cite{Halperin1982}. Since the bulk of a QH phase 
is gapped, its transport properties are controlled by the edge modes~\cite{Wen1990,Wen1992}.
The topological properties of the bulk QH phase guide the nature of these boundary modes~\cite{WenBook}.
For instance, conventional models suggest that integer and particle-like fractional phases support 
only {\it downstream} edge modes, while hole-conjugate phases may support modes with both (upstream and downstream)
chiralities~\cite{Wen91A,Wen91B,MacDonald_PRL_90,Johnson_PRL_91}.
Disorder-induced tunneling and intermode interactions further renormalise such counter-propagating edge modes and, 
in certain situations, may lead to the emergence of upstream neutral modes~\cite{KFP1994,KaneFisher_PRB_95,Yuval_AP_2017}.
Experimental signatures of these upstream neutrals have been observed in the $\nu = 2/3$~\cite{Bid2009,Bid2010,
Gurman2012,Yaron2012} and $\nu=5/2$ phases~\cite{Heiblum2021}, as well as in engineered geometries~\cite{Design1,Design2}.

However, recent transport experiments~\cite{Yacoby2012,Inoue2014,Heiblum2019} suggest that, even for 
the `simple' QH phases (such as the $\nu = 1$ or  $1/3$), the edge structure is much more intricate and may 
not be described by orthodox models. Exciting the $\nu = 1$ edge at a quantum point contact (QPC), Ref.~\cite{Yacoby2012}
observed an upstream flow of energy (but not charge). A similar experiment was performed for fractional QH phases
in the lowest Landau level in Ref.~\cite{Inoue2014}. They observed that 
partial transmission of charge current through a QPC is accompanied by upstream electric noise (with no net current) in several 
fractional QH phases (including Laughlin states). In a complementary study, Ref.~\cite{Heiblum2019} observed that the 
visibility of the interference pattern in an electronic Mach-Zehnder interferometer decreases as the filling factor 
($\nu$) is reduced from $2$ to $1$. Moreover, interference is fully suppressed for $\nu \leq 1$.
These experiments suggest that the standard picture of particle-like phases, involving one or more downstream modes, is incomplete. 
Instead these results point to the presence of additional counter-propagating modes, some of which may be charge-neutral.

\begin{figure*}[t]
  \centering
  \includegraphics[scale=0.25]{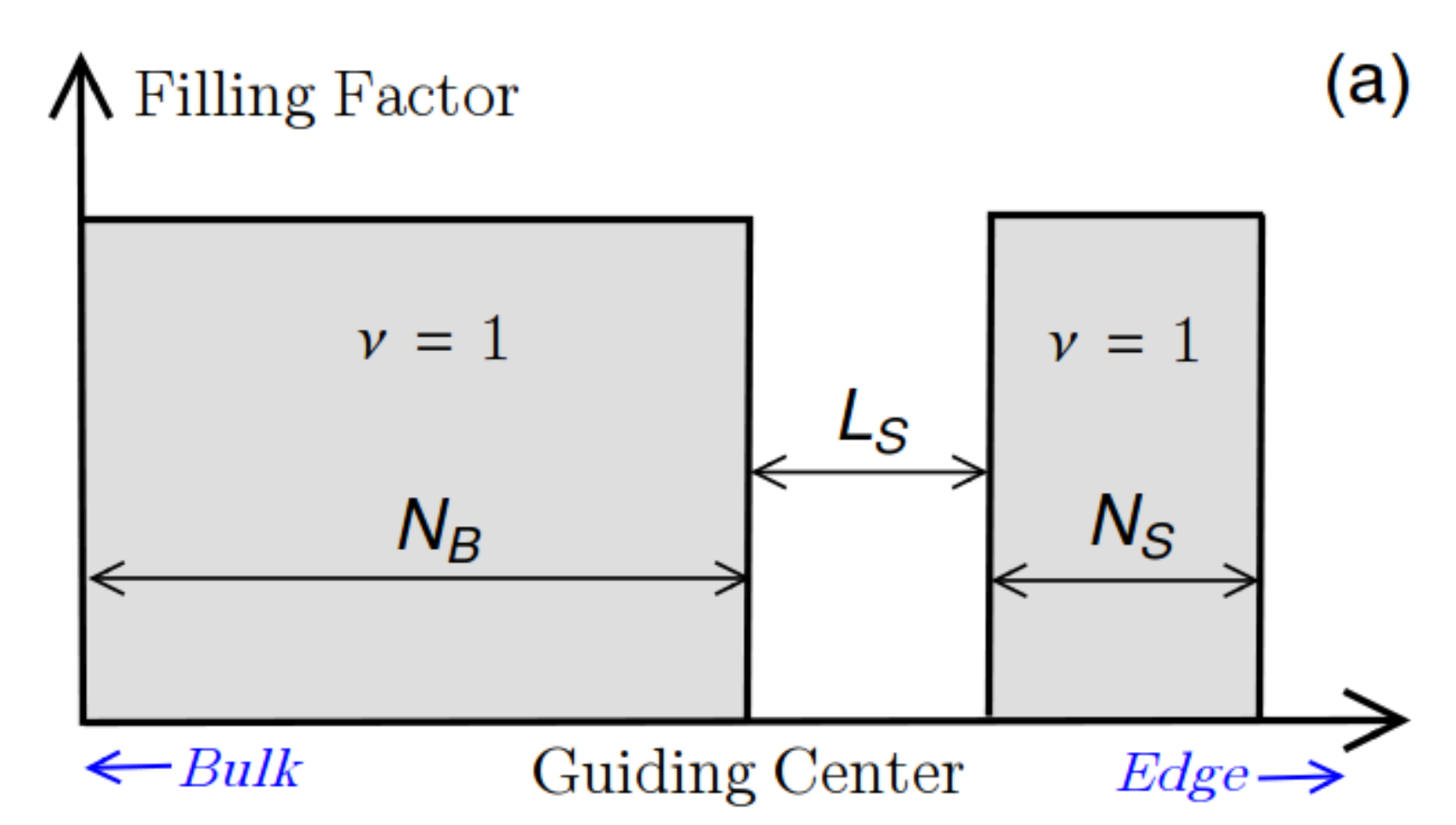}
  \includegraphics[scale=0.25]{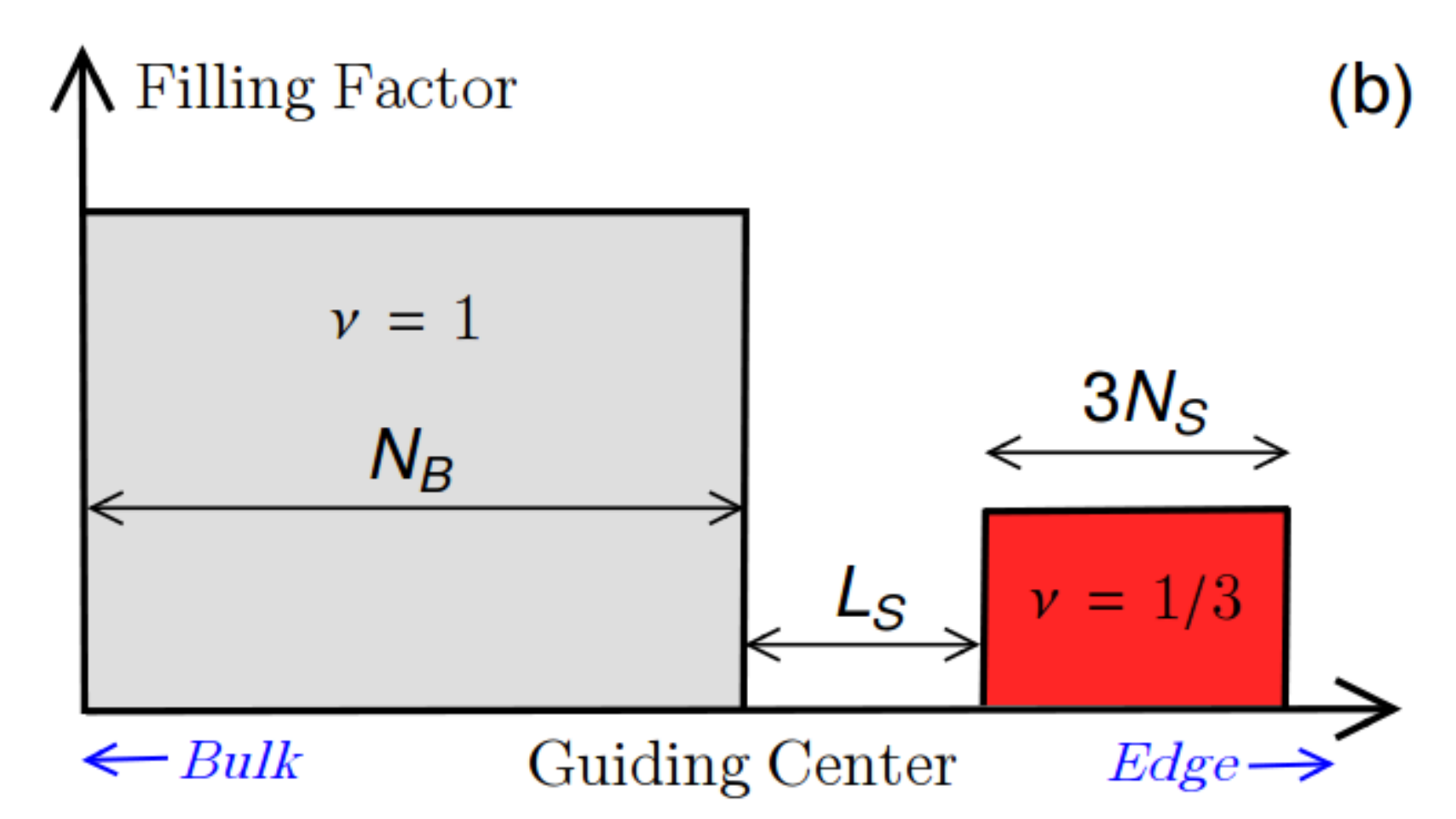}
  \includegraphics[scale=0.25]{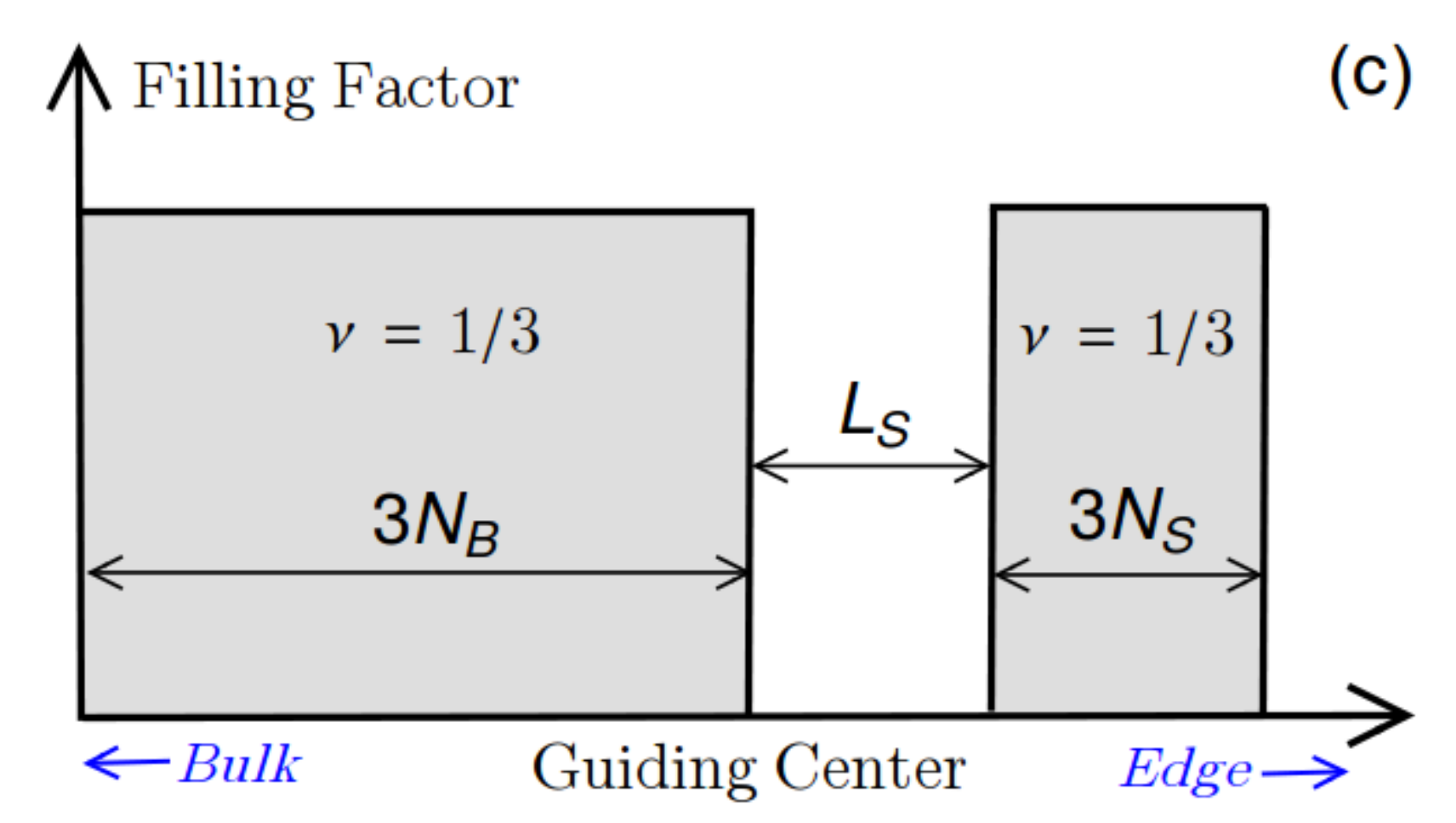}
  \includegraphics[scale=0.25]{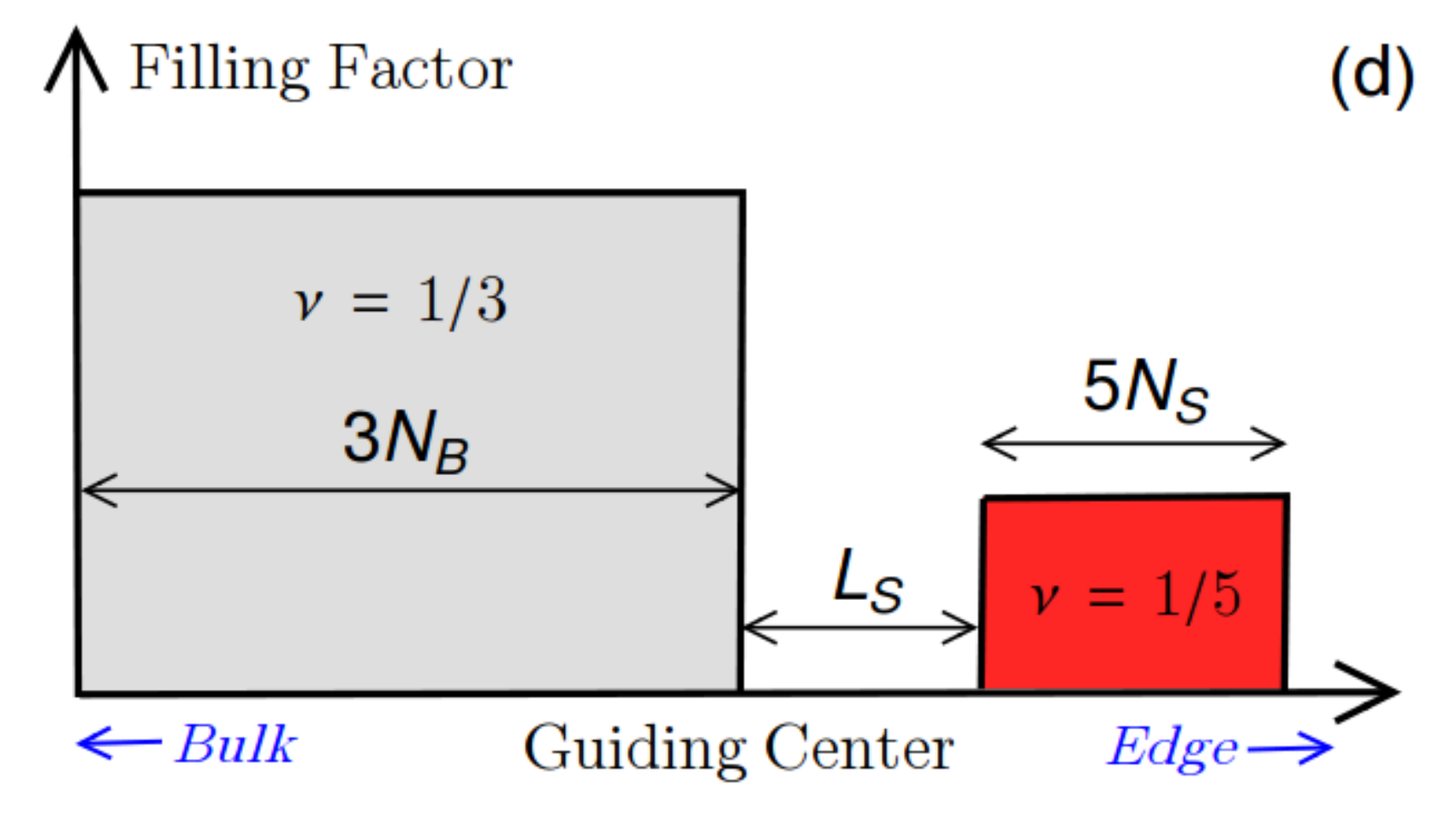}
  \caption{ A priori possible edge configurations for a bulk (a,b) $\nu = 1$ and (c,d) $\nu = 1/3$ phase. 
  For a sharp confining potential, a single QH droplet (with $\nu = 1$ or $1/3$) composed of $N_{B} + N_{S}$ electrons is
  expected. As the edge potential becomes smoother, an additional side strip (separated from the bulk by $L_{S}$ guiding centers) 
  composed of $N_{S}$ electrons is nucleated along the edge. The filling factor of the side strip may be the identical to 
  [as shown in (a,c)] or different from [as shown in (b,d)] the bulk filling factor.}
\end{figure*}

In the early 90s, it was realized that in the presence of a smooth confining potential at the boundary, electronic interactions
may induce quantum phase transitions at the edge (which leave the bulk unperturbed). Such edge transitions (or edge reconstructions) 
may occur in both integer~\cite{CSG1992,Dempsey1993,ChamonWen,Sondhi_PRL_96,FrancoBrey97,KunYangIQHS,Switching2017,
Ganpathy21,Karmakar2020,IQHS2020} and fractional~\cite{Meir93,MacDonald_JP_93,KunYang_2002,KunYang_2003,KunYang03,KunYang_2008,KunYang_2009,
Ganpathy_PRB_03,WMG_PRL_2013,Jain2014,Yang2021,FQHS2021,Liangdong21} QH phases, as well as in time-reversal-invariant topological 
insulators~\cite{Yuval2017,Rosenow2021}. The reconstructed edge structure may differ in terms of 
the number, order, or even the nature of the edge modes. Such phase transitions are driven by the 
competition between the electrostatic effects of a smooth confining potential and the exchange/correlation 
energy of an incompressible QH state. 
For sufficiently smooth potentials, this competition 
leads to nucleation of additional electronic strips (in QH phases) along the edge~\cite{Beltram2012,Thomas2014}. 
The nucleated side strips define additional pairs of counter-propagating chiral edge modes at their boundaries. 
Similarly to the edge of hole-conjugate states, intermode
interactions and disorder-induced tunneling among these additional and the original (topological) edge modes 
may lead to a subsequent renormalization, modifying their nature qualitatively. Such renormalization may even give rise to 
additional (non-topological) upstream neutral modes~\cite{WMG_PRL_2013}.

Here, we describe our recent attempts~\cite{IQHS2020,FQHS2021} to 
theoretically account for the experimental surprises described above, in terms of reconstruction and 
the subsequent renormalization of the edge for $\nu = 1$ and $1/3$ QH phases. Additionally, we present new analysis
of edge reconstruction of the $\nu = 2/5$ QH phase. 
The main challenge here is to determine the precise filling factor of the additional side strip nucleated at the edge
for a smooth confining potentials. 
An additional side strip of filling factor $\nu_{\text{strip}}$ defines counter-propagating modes of 
charge $\nu_{\text{strip}}$. Therefore, for $\nu_{\text{strip}} = \nu_{\text{bulk}}$ subsequent renormalization of 
the modes (due to disorder-induced tunneling) would lead to localization of a pair of counter-propagating modes and 
render transport experiments blind to the presence of reconstruction. On the other hand, for 
$\nu_{\text{strip}} < \nu_{\text{bulk}}$ subsequent renormalization 
would not induce localization, and may instead lead to the emergence of upstream neutral modes. 
Therefore, the experimental consequences of reconstruction crucially depend on the strip filling factor.

Figures~1, 2 depict some of the a priori possible configurations of the reconstructed edge at $\nu = 1,1/3$ and $2/5$. 
Here, we find the lowest energy state among these structures as a function of the slope of the confining potential through 
a variational analysis~\cite{Meir93,IQHS2020,FQHS2021}, which allows us to include a large number of electrons
while accounting for quantum correlations inherently present in QH states.
Specifically, we treat the strip-size ($N_{S}$) and separation ($L_{S}$) as variational
parameters. When the confining potential is sharp, we find the lowest energy state comprises a single QH droplet, i.e. no
edge reconstruction. On the other hand, for sufficiently smooth potentials, we find that edge reconstruction
leads to the emergence of a pair of additional counter-propagating gapless modes. Our results indicate that, in all
three phases, the gapless modes of the reconstructed edge, and their subsequent renormalization leading to the 
emergence of neutral modes, may account for the experimental results reported in Refs.~\cite{Yacoby2012,Inoue2014,Heiblum2019}.
We also analyze additional experimental consequences of edge reconstruction, such as in two-terminal transport.

\section{Model Hamiltonian and Variational Analysis}

{\it Here, we provide details of the theoretical model used to study the QH edge. We also describe the 
variational analysis employed to find the lowest energy edge configuration as a function of the
slope of the confining potential.}

\begin{figure*}[t]
  \centering
  \includegraphics[scale=0.25]{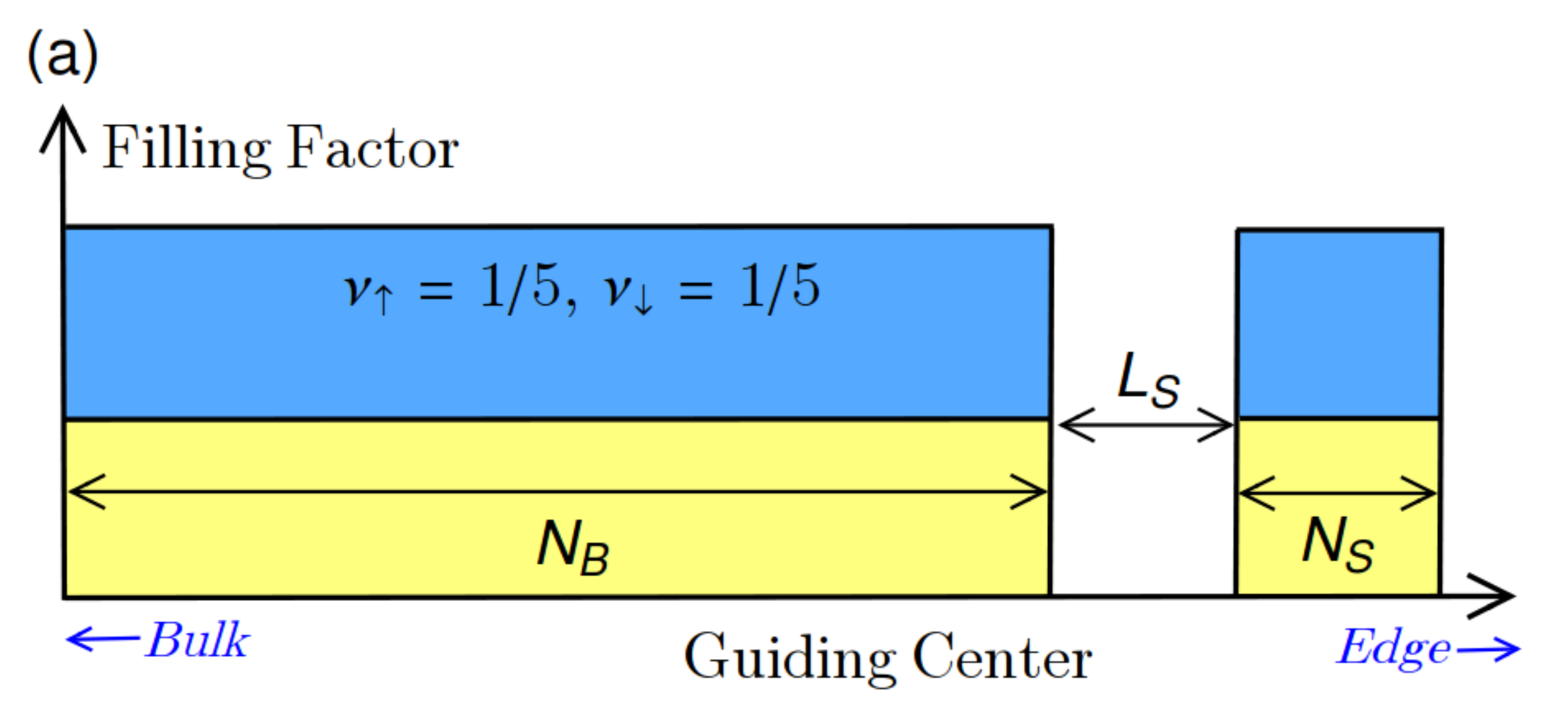}
  \includegraphics[scale=0.25]{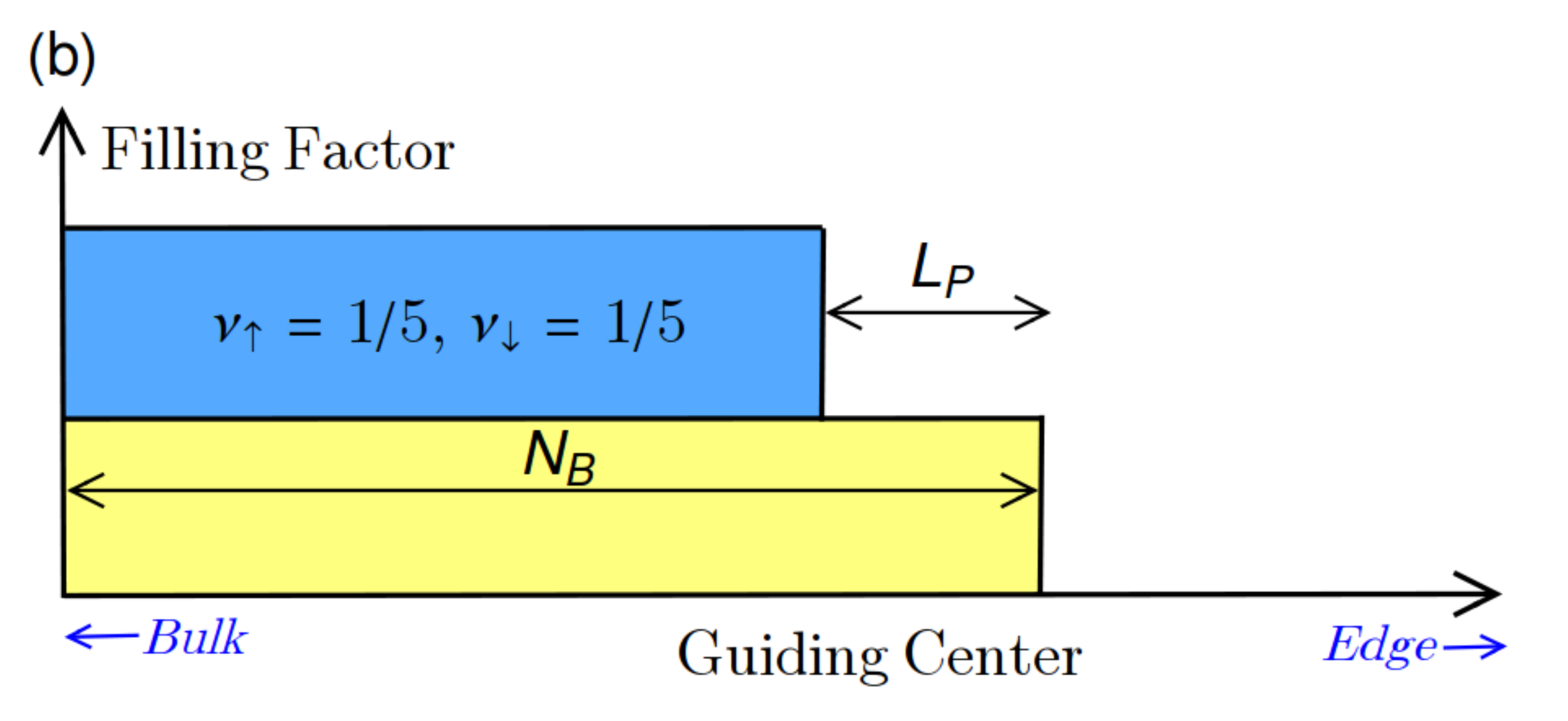}
  \includegraphics[scale=0.25]{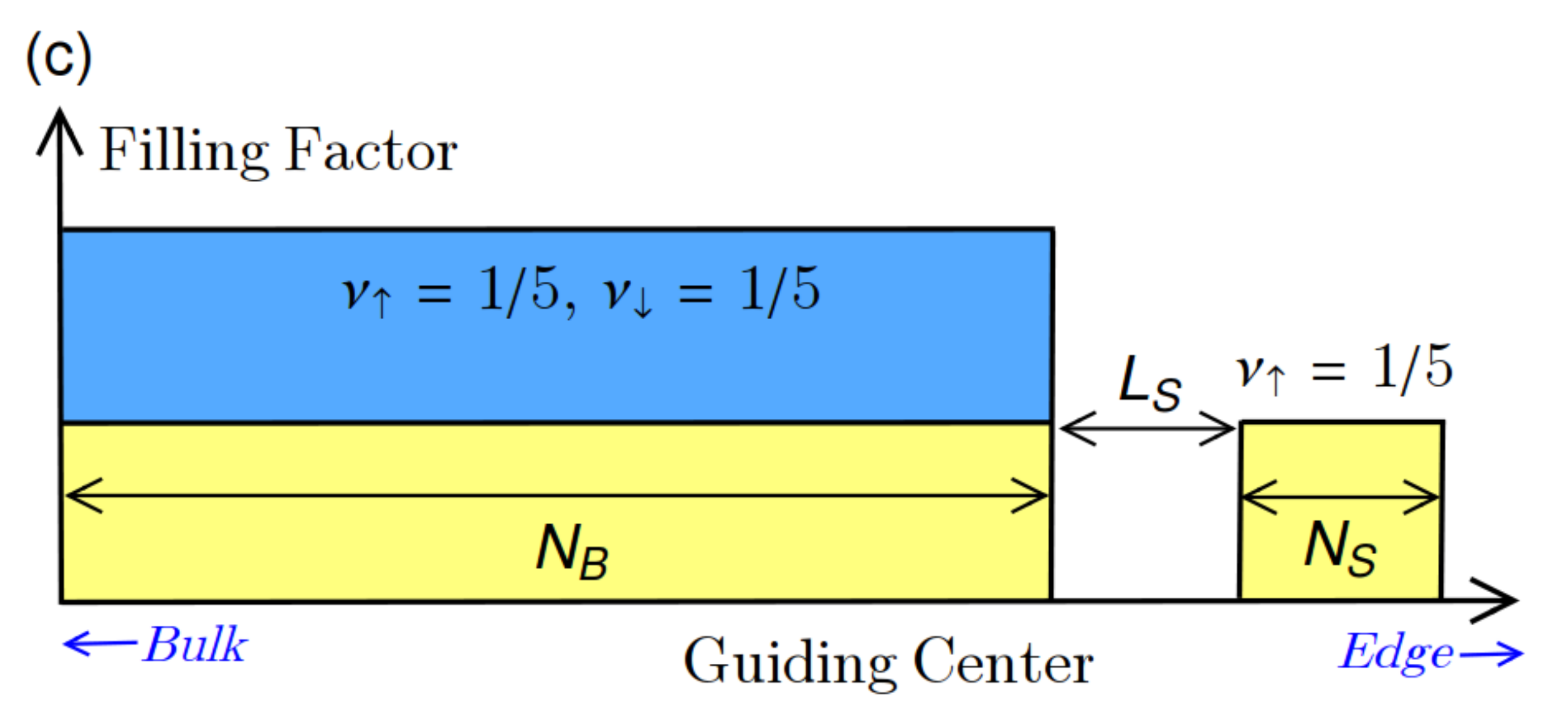}
  \includegraphics[scale=0.25]{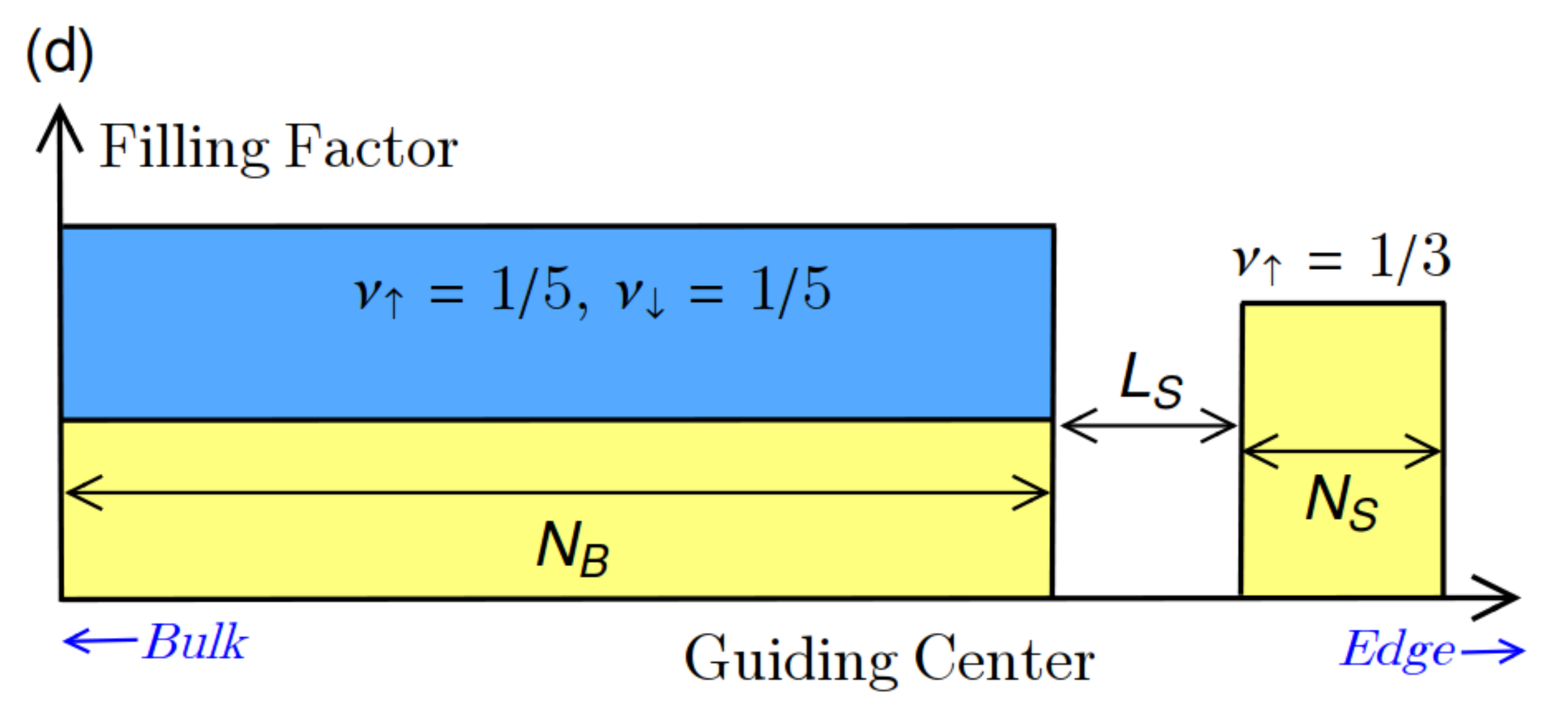}
  \caption{A priori possible structures at the reconstructed edge for a (spin-singlet) $\nu = 2/5$ phase. 
  The blue and yellow colors correspond to the two spin-polarizations of electrons in the lowest Landau level. 
  Panel (a) depicts a spin-unpolarized edge configuration, while panels (b-d) depict structures with finite 
  magnetization at the edge. Such spontaneous magnetization may arise without an additional stripe at the edge [as
  depicted in panel (b)] or due to the formation of such a stripe [panels (c,d)].}
\end{figure*}

\subsection{Basic Setup}

We analyze the QH edge in the disk geometry, which is convenient due to the presence of a single boundary even for finite systems. 
We employ a rotationally symmetric gauge, $e \vec{A} /\hbar = (-y/2\ell^{2}, x/2\ell^{2})$, where $\ell = \sqrt{\hbar/eB}$ is the magnetic length.
The rotational invariance allows the single-particle states to be labelled by eigenvalues
of the angular momentum ($\hat{L}$). We denote the states in the lowest Landau level (LLL) as $\phi_{m}$ with $m = 0, 1, 2, \dots$.
The corresponding wavefunction is $\phi_{m} (\vec{r}\,)
= \left( r / \ell \right)^{m} e^{-im\theta_{\text{r}}} e^{-\left(\frac{r}{2\ell}\right)^2} / \sqrt{2^{m+1} \pi m! \ell^2} $,
where $re^{-i \theta_{\text{r}}} = x - iy$ is the electronic position. The state $\phi_{m}$ is
strongly localized around $r = \sqrt{2 m} \ell$ and has angular momentum $\hbar m$.

In the LLL, the dynamics may be described by the 
Hamiltonian $H = H_{\text{ee}} + H_{\text{c}}$,
where $H_{\text{ee}}$ is the two-body electronic repulsion and $H_{\text{c}}$ is the one-body confining potential 
(also assumed to be circularly symmetric).  
Note that $H$ commutes with $\hat{L}$. Therefore, the many-body states
may be labelled by the total angular momentum.
Defining $E_{c} = e^2/\epsilon_0 \ell$ as the Coulomb energy scale and $c_{m \sigma}$ as the annihilation operator
corresponding to $\phi_{m}$ with spin state $\sigma=\ua/\da$ (along $s_{z}$), we have,
\begin{align}
  H_{\text{ee}} &= \frac{E_{\text{c}}}{2} \sum_{i \neq j} \frac{\ell}{|\vec{r}_{i} - \vec{r}_{j}|} \\ \nonumber
  &\equiv \frac{E_{\text{c}}}{2} \sum_{\{m,\sigma\},n} V_{m_1 m_2 ; n}^{ee}
  c_{m_1 + n \sigma_{1}}^{\dagger} c_{m_2 \sigma_{2}}^{\dagger} c_{m_2 + n \sigma_{2}} c_{m_1 \sigma_{1}}, \\
  H_{\text{c}} &= \sum_{m, \sigma} V_{m}^{\text{c}} \, \, c_{m \sigma}^{\dagger} c_{m \sigma}. \label{eq:H2}
\end{align}

The confining potential 
is modelled using a positively charged background disk (with radius $R$, charge density $\rho_{\text{bg}}$) 
separated from the electron gas by a distance $d$ along the magnetic field~\cite{KunYangIQHS,KunYang_2002,KunYang_2003}. 
The parameters $R$ and $\rho_{\text{bg}}$ are chosen such that overall charge neutrality is maintained. 
The electrostatic potential of this disk (in the plane of the electrons) is, 
\begin{align}
  V_{c} (r) = \int_{0}^{R} dr^{\prime} \int_{0}^{2\pi} d \theta \frac{E_{c} \rho_{\text{bg}}}{\sqrt{d^2 + r^2 + {r^{\prime}}^2 -
  2 r^{\prime} r \cos \theta}}.
\end{align}
Then $V_{m}^{\text{c}}$ in Eq.~(\ref{eq:H2}) are the matrix elements of $V_{c}(r)$.
Note that the smoothness of this confining potential is controlled by the distance $d$ (the tuning parameter in our analysis). 
Specifically, the edge potential is quite sharp for $d \sim 0$, and becomes smoother as $d$ increases. 

We note that edge reconstruction of both integer and fractional QH phases has been considered in previous works. 
These studies employed unbiased methods, such as exact diagonalization (ED)~\cite{KunYangIQHS,KunYang_2002,KunYang_2003,
KunYang_2008,KunYang_2009} and density matrix renormalization group (DMRG)~\cite{Liangdong21,DMRG2021}. 
However, the precise filling factor at the edge cannot be obtained in ED due to its inherent limitation to small
system sizes. By contrast, DMRG overcomes the size limitations of ED but works best for one-dimensional 
systems and its applicability to the problem of edge reconstruction is not clear. For these reasons, here we 
employ a variational method to study the edge~\cite{Meir93}. Our method is capable of predicting the precise 
filling factor of the edge, and is not limited to a small system size. Moreover, such methods have been used 
extensively to study various aspects of QH phases~\cite{JainCF} and their applicability is well established.

\subsection{Variational Analysis}

We consider several variational classes describing a priori possible edge structures for the $\nu = 1, 1/3$ and $2/5$ 
QH phases. All the classes considered here represent product states of a bulk QH droplet composed of $N_{B}$ electrons, 
and a single edge strip composed of $N_{S}$ electrons. The edge strip is separated from the bulk by $L_{S}$ guiding centers. 
In our analysis, the total number of electrons ($N_{B} + N_{S}$) is kept fixed. Therefore, the states in any of the classes 
may be parameterized by $N_{S}$ and $L_{S}$. For a given bulk QH phase, each variational class corresponds to fixed filling
factors for the bulk and the edge strip. 
Figures 1 and 2 depict the various classes of variational states considered in this work.

\begin{figure}[t]
  \centering
  \includegraphics[width=0.95\columnwidth]{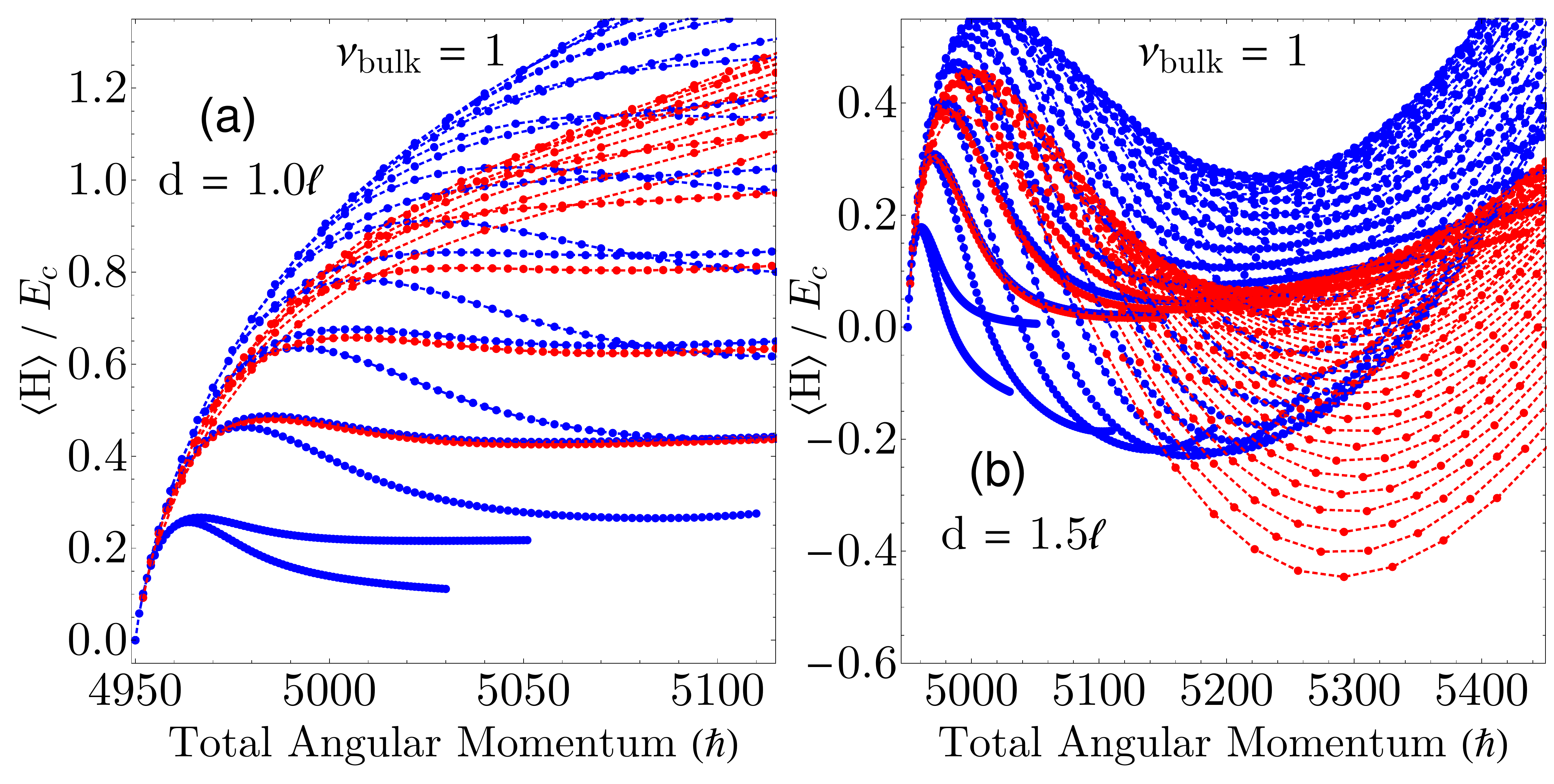}
  \caption{ Results of the variational analysis using $100$ electrons with bulk filling factor $\nu = 1$. 
  The blue (red) dots show the total energy
  of the variational states with a $\nu = 1$ integer ($\nu = \frac{1}{3}$ fractional)
  side-strip as a function of the total angular momentum for a (a) sharp ($d = \ell$)
  and (b) smooth ($d = 1.5 \ell$) confining potential.
  Each curve corresponds to states with the same separation between the bulk and side-strip ($L_{S}$)
  but with different number of electrons in the side-strip ($N_{S}$). The curves shown here correspond
  to $L_{S}$ varying from $0$ to $30$ guiding centers.
  The energy of the unreconstructed state has been subtracted to make comparison easier.
  (a) For sharp edges $(d < 1.3 \ell)$ the ground state is the one with minimum angular momentum,
  implying no edge reconstruction. (b) For smooth edges $(d > 1.3 \ell)$ the ground state shifts to
  a higher angular momentum sector, implying that the electronic disk expands and the edge
  undergoes reconstruction. The minimum energy state lies on the curve corresponding to $L_{S} = 0$.
  Panel (b) shows that a fractional reconstruction is energetically favorable to
  an integer reconstruction. }
\end{figure}

The $\nu = 1$ and $1/3$ phases are assumed to be fully spin-polarized (cf. Fig.~1). Therefore we consider spinless electrons in these cases. 
Then the bulk $\nu = 1$ phase represents a Slater determinant of $N_{B}$ electrons, occupying all the guiding centers from $m = 0$ to 
$m = N_{B} - 1$. The bulk $\nu = 1/3$ phase is represented by the $\nu = 1/3$ Laughlin state. 
The Laughlin wavefunction corresponding to $\nu = 1/m_{B}$ is~\cite{Laughlin83,JainCF},
\begin{align}
  \Psi_{\frac{1}{m_{B}}, N_{B}} = \prod_{j > i} \bigg[ \big( z_i - z_j \big)^{m_{B}} \bigg] e^{-\frac{1}{4} \sum_{i} |z_{i}|^2}. 
\end{align}
Here $z_{j} = (x_{j} - iy_{j})/\ell$ is the position of the $j^{th}$ electron. Next, the edge strip comprising the $\nu = 1$ phase [Fig.~1(a)]
may also be represented as a Slater determinant of $N_{S}$ electrons. On the other hand, the edge strip comprising $\nu = 1/3, 1/5$ phases
[Figs. 1(b-c)] are described through a $\nu = 1/m_{S}$ Laughlin state ($m_{S} = 3, 5$) 
with $M_{S}$ quasiholes at the origin. The separation of the bulk and edge strip ($L_{S}$) is given by, 
$L_{S} = (M_{S} - 1) - \nu_{\text{bulk}} (N_{B} - 1) $. 
The corresponding (unnormalized) wavefunction is, 
\begin{align}
  \Psi_{\frac{1}{m_{S}}, N_{S}, M_{S}} = \prod_{i=1}^{N_{S}} \bigg[ z_{i}^{M_{S}} \,
  \prod_{j > i} \big( z_i - z_j \big)^{m_{S}} \bigg] e^{-\frac{1}{4} \sum_{i} |z_{i}|^2}. 
\end{align}

In this work, we focus on the spin-unpolarized $\nu = 2/5$ phase, which may be described as the product state of two 
copies (one for each spin) of the $\nu = 1/5$ Laughlin phase, i.e. $\Psi_{\frac{2}{5}, N_{B}} = 
\Psi_{\frac{1}{5}, N_{B}/2,\ua} \otimes \Psi_{\frac{1}{5}, N_{B}/2,\da}$. The reconstructed edge in this phase could be 
`simple' and identical to the bulk [as shown in Fig.~2(a)], or due to the additional spin degree of freedom, may be 
spontaneously spin-polarized [see Figs.~2(b-d)]. Interestingly, the latter possibility may occur even without the nucleation of an additional 
edge stripe [see Fig.~2(b)] (this is analogous to the edge structure described for the $\nu = 2$ phase in Ref.~\cite{Dempsey1993}). 
All these configurations may be described through product states of appropriate Laughlin states, as mentioned above. 

For Slater determinants, the energy ($\langle H \rangle$) of the variational states may be evaluated trivially 
given the matrix elements of the Coulomb interaction and the confining potential~\cite{IQHS2020}. On the other hand,
for Laughlin states these may be evaluated using standard classical Monte-Carlo 
techniques~\cite{JainCF,Metropolis53,Laughlin83,MacDonald93,FQHS2021}. 
In our analysis, we evaluate the energy of the states in each variational class as a function of $d$, which controls
the slope of the confining potential. The ground state for each QH phase, and the precise structure of the edge, is then 
found by comparing the energies of the states in the different classes. Note that the unreconstructed state (without an 
additional edge strip) is included in all classes (corresponding to $N_{S} = 0 = L_{P}$). 
It is the lowest energy state for sharp confining potentials ($d \sim 0$). By contrast, the ground state supports an additional
edge strip (finite $N_{S}$, $L_{S}$ or $L_{P}$) for smoother potentials. Finally, the structure of the edge in the ground state
uniquely determines the number and nature of the low-energy chiral modes.

\section{Variational Results}

{\it This section presents the results of our analysis of the edge structure for the $\nu = 1, 1/3$ and $2/5$ QH phases. 
In all cases, we find that edge reconstruction may lead to the emergence of side stripes with filling factor different
from that of the bulk QH phase. } 

\begin{figure*}[t]
  \centering
  \includegraphics[width=0.95\textwidth]{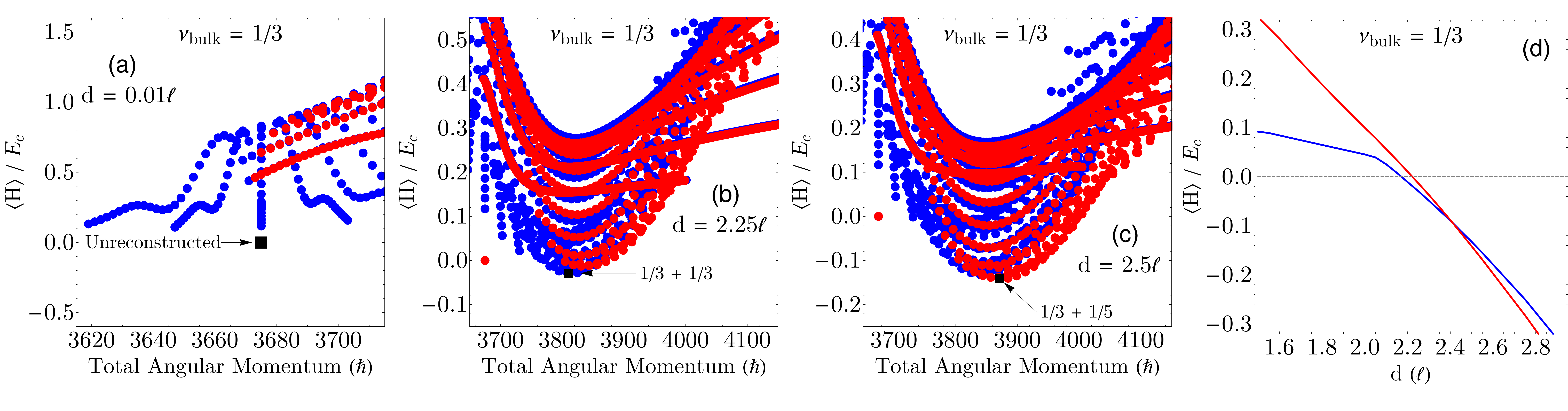}
  \caption{ Results of the variational calculations with 50 electrons with bulk filling factor $\nu = 1/3$. 
  (a)-(c) The energy ($\langle H \rangle$) of the 
  states in the two variational classes as a function of the 
  total angular momentum at (a) sharp ($d = 0.01 \ell$), (b) 
  moderately smooth ($d = 2.25 \ell$), and (c) very smooth 
  ($d = 2.50 \ell$) confining potentials.
  In all cases, the energy of the unreconstructed state ($\langle H \rangle_{\text{ur}}$) has been
  subtracted. The blue (red) circles show energy of states with a side strip of $\nu = 1/3$ ($\nu = 1/5$). 
  The black square marks the state with the 
  lowest energy. (d) The variation of the lowest possible energy in the two variational classes 
  with the smoothness of the confining potential (parameterized by $d/\ell$). The blue 
  (red) line corresponds to states with a side strip of $\nu = 1/3$ ($\nu = 1/5$). 
  As expected, for sharp edges the ground state is the one with $N_{S} = 0$, corresponding
  to the unreconstructed $\nu = 1/3$ state with angular momentum $3675 \hbar$. This state supports a single 
  chiral $e/3$ mode. For moderately smooth potentials ($ 2.17 < d/\ell < 2.42$), an additional strip of 
  $\nu = 1/3$ is generated at the edge, which gives rise to an extra pair of counter-propagating
  $e/3$ modes. For very smooth potentials ($d > 2.42 \ell$) the additional strip has the filling
  factor $1/5$. This second reconstructed state supports a counterpropagating pair of $e/5$ modes
  in addition to the chiral $e/3$ mode arising from the bulk. }
\end{figure*}

Figure~3 shows the total energies for the two classes of variational states corresponding to $\nu = 1$ as a function of
the total angular momentum at different confining potentials (controlled by $d$). A total of 100 particles were used for
these results. The blue dots correspond to 
integer edges [Fig.~1(a)] while the red dots correspond to the fractional edges [Fig.~1(b)]. For a sharp confining 
potential [$d < 1.2\ell$, Fig.~3(a)] the lowest energy state is the one with the minimal angular momentum 
(in this case $4950\hbar$). This corresponds to the unreconstructed $\nu = 1$ state with a single chiral edge mode. 
For smoother potentials [$d > 1.3\ell$, Fig.~3(b)], the lowest energy state has a much larger angular momentum
($5256 \hbar$ for $d=1.5\ell$ with $N_{S} = 18$ and $L_{S} = 0$) than the compact state. 
Note that the states with a fractional edge are found to have a lower energy than the states with an integer edge
\textit{whenever reconstruction is favored}. We have verified that our results do not depend on the detailed form
of the confining potential~\cite{IQHS2020}. 
The fractionally reconstructed edge [Fig.~1(b)] supports a downstream 
$e^{*} = 1$ mode (originating from the bulk) in addition to a counter-propagating pair of $e^{*} = 1/3$ modes 
arising from the side strip. Therefore, our results imply that {\it fractionally} charged chiral edge modes may 
exist even at the edge of bulk integer QH phases.

\begin{figure}[b]
  \centering
  \includegraphics[scale=0.25]{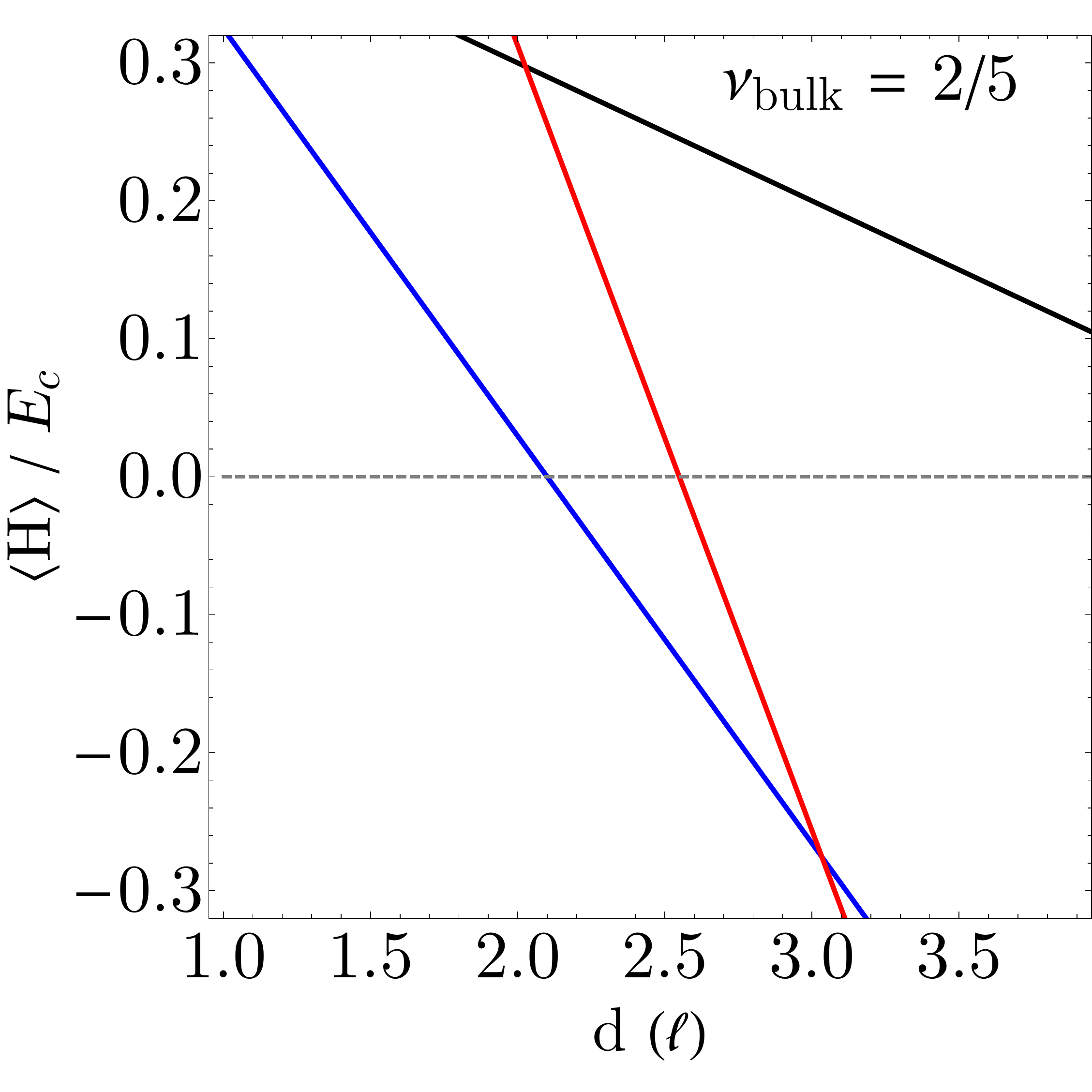}
  \caption{ Results of the variational calculations with 40 electrons with bulk filling factor $\nu = 2/5$. 
  The curves show the variation of the lowest possible energy in the different variational classes as a 
  function of the smoothness of the confining potential. The energy of unreconstructed state has been subtracted for ease of comparison. 
  The black curve corresponds to spin-unpolarized edge, while the blue (red) curves correspond to spin-polarized edge structures 
  without (with) an additional edge stripe. The red curve corresponds to an edge stripe in the $\nu = 1/3$ phase. }
\end{figure}

Figure~4 depicts the total energies of the states in the two classes corresponding to $\nu = 1/3$, classified by their 
angular momentum, for several values of $d$. These results correspond to a total of 50 particles. The blue (red) dots in Figs.~4(a-c)
correspond to edges with a side strip of filling factor $1/3$ ($1/5$). The black square marks the lowest 
energy state. In each case, we have subtracted the energy of the unreconstructed state ($N_{S} = 0$) to 
make the comparison easier. For a sharp confining potential [$d \lesssim 2.1\ell$, Fig.~4(a)] the standard
Laughlin state, with no additional side strip, has the lowest energy (as expected). 
Such a state clearly has a single chiral $e/3$ mode at the edge.
For smoother potentials [$d \gtrsim 2.1\ell$, Figs.~4(b-c)], the lowest energy state comprises an 
additional side strip ($N_{S} > 0$). This side strip may have filling factor $1/3$ [Fig.~4(b)] for moderate slope 
of the confining potential ($N_{S} = 15$, $L_{S} = 11$ for $d = 2.25 \ell$) or $1/5$ [Fig.~4(c)] for 
very shallow slope of the potential ($N_{S} = 14$, $L_{S} = 3$ for $d = 2.5 \ell$). 
Figure~4(d) shows the variation of the lowest possible energy in the two classes with the slope of the
confining potential. Evidently, the filling factor of the side strip is $1/3$ in the range 
$2.17 \ell < d < 2.42 \ell$, and switches to $1/5$ for larger values of $d$. 
Hence, our analysis of the $\nu = 1/3$ edge suggests that upon reconstruction, it may support 
in addition to the single $e^{*} = 1/3$ mode arising from the bulk, a pair of counter-propagating $e^{*} = 1/3$ or 
(notably) $1/5$ modes.

Figure~5 presents the lowest possible energy in the classes corresponding to $\nu = 2/5$ as a function of the slope of
the confining potential. These are results for a total (including both spins) of 40 particles. The black line corresponds
to the structure shown in Fig.~2(a) with a finite $N_{S}$ (note that $N_{S} = 0$ corresponds to the unreconstructed state). 
Our analysis suggests that such a reconstruction is not energetically favorable for any slope of the confining potential. 
The blue line corresponds to reconstruction without an additional strip [Fig.~2(b)]. Clearly, this edge configuration is 
favorable (compared to the unreconstructed state) for $d > 2.0 \ell$. The emergence of a spontaneous spin-polarization at the edge 
through a redistribution of the particles within the bulk (as opposed to the formation of a separate stripe) is analogous to 
the results of Ref.~\cite{Dempsey1993} for the bulk $\nu = 2$ state. Such a reconstruction does {\it not} lead to the emergence of
new chiral modes. Rather, it only increases the spatial separation between the two bare (spin-polarized) $e^{*} = 1/5$ modes supported
by the bulk state. However, our analysis suggests that for even smoother confining potentials ($d > 3.1 \ell$), a separate 
edge stripe with $\nu = 1/3$ [Fig.~2(d)] is more favorable energetically (the red curve in Fig.~5 shows the lowest possible 
energy of this class). Such an edge structure has finite edge magnetization and supports an (additional) pair of counter-propagating 
$e^{*} = 1/3$ modes. Our results indicate that the structure shown in Fig.~2(c) is not energetically favorable in any range of 
$d$. For this reason, we do not show the energy of this class in Fig.~5. We thus conclude that for sufficiently smooth confining
potentials, the spin-unpolarized $\nu = 2/5$ state may support at its edge, a pair of (spin-polarized) counter-propagating 
$e^{*} = 1/3$ modes in addition to the pair of downstream $1/5$ modes of both spins.

\section{Experimental Manifestations of Edge Reconstruction}

{\it The various configurations of the reconstructed edge found in our analysis may be uniquely identified in carefully
designed transport experiments. Here, we focus on the behavior of the two-terminal conductance as a function of the sample 
length, and the manifestations of upstream neutral modes, which may emerge due to further renormalization of the edge modes. }

\subsection{Two-Terminal Conductance}

Edge reconstruction is expected to have very clear consequences for the (electric) two terminal conductance 
($\gt$) as a function of the length of the edge ($L$). In a two-terminal setup, in the absence of edge equilibration,
the chiral channels exiting the source contact are biased with respect to the modes entering it.
The presence of impurities and potential disorder generates random tunneling between the chiral modes
at the edge, which may facilitate intermode equilibration over a characteristic length $\ell_{\text{eq}}$~\cite{Nosiglia2018,Gornyi21}. 
Therefore, we may expect that $\gt$ varies as a function of $L$ over the equilibration length scale $\ell_{\text{eq}}$.  
For $L \gg \ell_{\text{eq}}$, assuming full intermode equilibration, the two-terminal conductance is
$\gt = \nu_{\text{bulk}} \times e^{2}/h$ irrespective of the slope of the confining potential, reflecting the 
topological order of the bulk.

\begin{figure}[t]
  \centering
  \includegraphics[width=0.98\columnwidth]{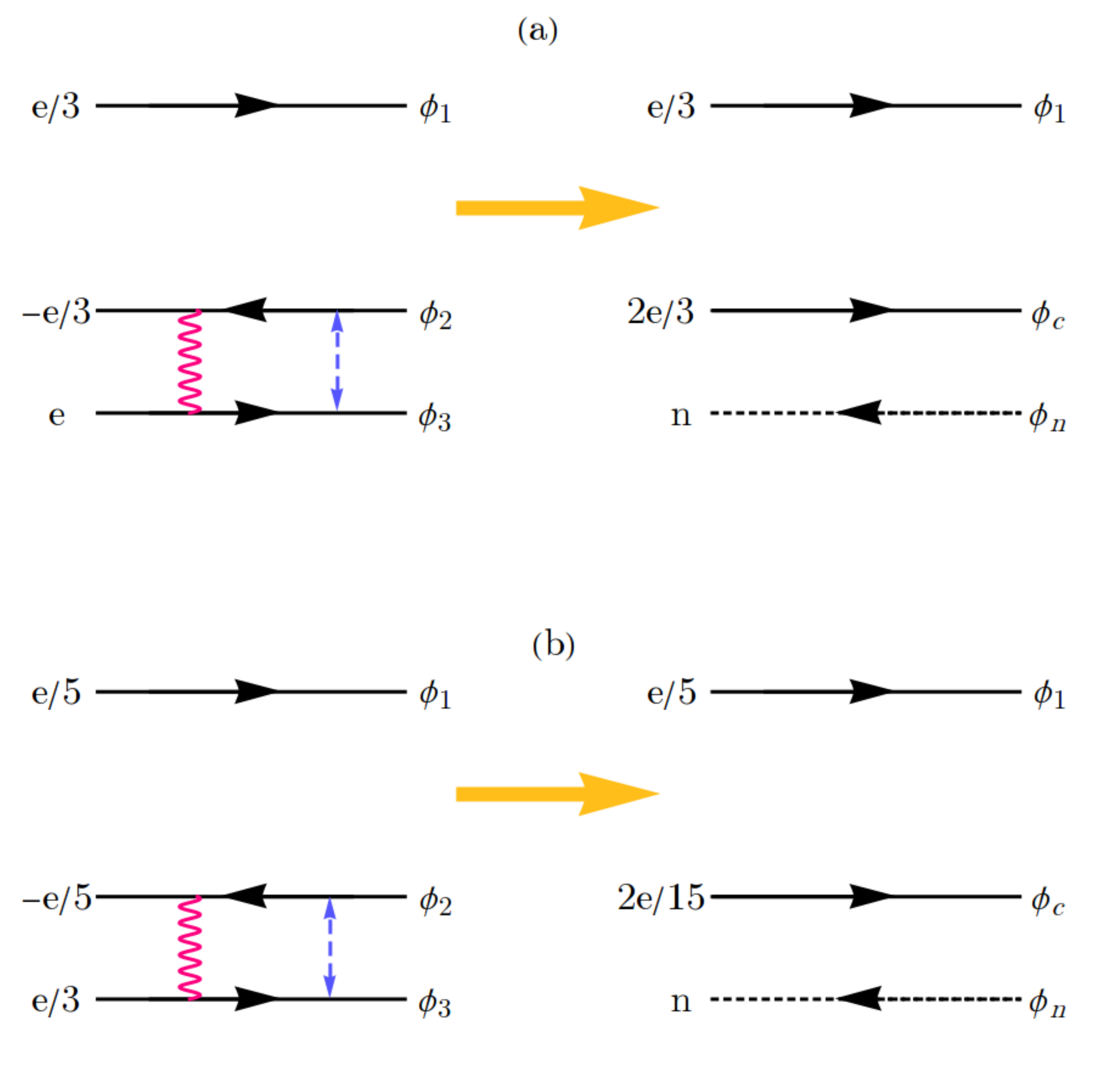}
  \caption{ For the edge structures in Figs.~1(b,d), the bare edge modes ($\phi_{1,2,3}$) are renormalized 
  by intermode interactions (represented by the red wavy line) and disorder-induced 
  electron tunneling (represented by the blue dashed line). Such renormalization may lead to emergence of a 
  downstream charge ($\phi_{c}$) and an upstream neutral mode ($\phi_{n}$). 
  In both cases, the outermost mode is assumed to be decoupled from the inner two modes, since our variational
  analysis indicates that, for smooth confining potentials, the distance between the outer mode and the two 
  inner modes is much larger than the distance between the two inner modes.}
\end{figure}

The $L \ll \ell_{\text{eq}}$ regime is more interesting, since the two terminal conductance is sensitive
to the detailed structure of the edge in absence of intermode equilibration. For the unreconstructed edge 
(in the case of a sharp confining potential), $\gt = \nu_{\text{bulk}} \times e^{2}/h$ for all values of $L$. This is because
the unreconstructed edge supports only downstream mode (for the three phases considered here), rendering the
notion of equilibration irrelevant. 

For reconstructed edges the additional pair of counter-propagating modes may also contribute to $\gt$. 
For the bulk $\nu = 1$ phase, the side stripe has filling factor $\nu = 1/3$. Then $\gt = 5/3 \times e^{2}/h$ in 
very short samples. Note that the coefficient $5/3$ uniquely determines the filling factor of the edge stripe. 
Hence, this is a {\it smoking gun} signature of our predicted edge structure. The reconstructed edge of 
the $\nu = 1/3$ phase, may support additional modes with $e^{*} = 1/3$ or $1/5$ depending on the slope of the confining potential. 
Evidently, $\gt = 1 \times e^{2}/h$ ($11/15 \times e^{2}/h$) in the former (latter) case. 
Finally, for the $\nu = 2/5$ phase, $\gt$ is sensitive only to the `second' reconstruction involving the formation of 
an additional $\nu = 1/3$ stripe. In this case, $\gt$ increases to $16/15 \times e^{2}/h$. We note that the length 
dependence of the two-terminal conductance has been reported for other filling factors~\cite{Lafont2019}.

\subsection{Emergent Non-Topological Neutral Modes}

In the previous section, we relied on our variational analysis of the ground state in order to discern the nature of the chiral modes 
at the reconstructed edge. However, intermode interactions and disorder-induced tunnelling among these chirals may lead a subsequent 
renormalization of the bare edge modes. Such renormalization would lead to localization for identical (same $e^{*}$) counter-propagating 
modes. By contrast, counter-propagating modes of unequal charges (arising from QH regions of different filling factor) would be 
renormalized to two new effective modes of (in general, non-universal) charges $e^{*}_{\ua}$ and $e^{*}_{\da}$ (here, $\ua/\da$ denotes the 
direction of propagation: upstream/downstream)~\cite{KFP1994,Yuval_AP_2017}. Interestingly, in some cases $e^{*}_{\ua}$ may be zero leading 
to the emergence of gapless upstream neutral modes. 

As explained previously, our variational analysis suggests that (for sufficiently smooth potentials) the filling factor of the 
additional side strip is not equal to the bulk filling factor, implying that chiral modes with differing $e^{*}$ may be 
supported at the reconstructed edge. Our results also indicate that as the confining potential becomes shallower ($d$ increases), 
the width of the edge stripe ($N_{S}$) increases much faster than its separation from the bulk ($L_{S}$). Hence, for very smooth 
confining potentials the outermost chiral mode couples very weakly to the inner pair of counter-propagating modes. Over sufficiently short length 
scales, we may assume that the outermost mode is completely decoupled from the other two. In this case, mode renormalization of the
inner pair of counter-propagating chirals could lead to upstream neutral modes ($\phi_{n}$ in Fig.~6). Note that for simplicity, 
we only focus on the fully spin-polarized cases of $\nu = 1$ and $\nu = 1/3$ phases in this section. 

The emergent neutral mode $\phi_{n}$ supports chiral flow of heat without an accompanying charge current, and hence has 
several unique manifestations in transport experiments. Such an upstream heat current was reported 
in Ref.~\cite{Yacoby2012} for the $\nu = 1$ phase. A biased neutral mode may also lead to generation of shot noise, despite the
absence of a net charge current, due to the formation of quasiparticle-quasihole pairs~\cite{Park2019,Spanslatt2019,Spanslatt2020}. 
Such observations were reported in Refs.~\cite{Inoue2014,Sabo2017,Heiblum2019} for various QH phases (including particle-like fractions). 
Additionally, the presence of upstream neutral modes may lead to the generation of shot noise on the (intermediate) conductance plateaus in
the transmission of a quantum point contact. Interestingly, under certain situations, the Fano factor of this noise may be quantized and
equal to the bulk filling factor $\nu_{\text{bulk}}$ instead of the quasiparticle charge~\cite{Heiblum2019,Cohen2019,Jinhong2020,Biswas2021}.
A complementary signature of upstream neutrals is the suppression of visibility of anyonic interference in electronic Mach-Zehnder 
setups~\cite{Moshe_PRL_2016}. This is in accordance with the observations of Ref.~\cite{Heiblum2019} for QH phases with $\nu \leq 1$.

\section{Conclusions}

We have employed variational analysis to study edge reconstruction that at the boundary of prototypical 
particle-like QH phases ($\nu = 1, 1/3$ and $2/5$).  
We have found that, in each case, edge reconstruction leads to the formation of an additional side strip, and that for
sufficiently smooth potentials, the filling fraction of this side strip may be different from the bulk filling factor. 
Such a reconstruction has clear signatures in transport. We have pointed out some of these 
consequences related to the two-terminal conductance and the emergence of upstream neutral modes.  \\

\section*{Acknowledgments}

We acknowledge illuminating discussions with Jinhong Park and Moty Heiblum. 
M.G. was supported by the Israel Science Foundation (ISF) and the Directorate for Defense Research and 
Development (DDR\&D) grant No. 3427/21 and by the US-Israel Binational Science Foundation (BSF) Grants No. 
2016224 and 2020072. Y.G. was supported by CRC 183 (project C01), the Minerva Foundation, DFG Grant No. 
RO 2247/11-1, MI 658/10-2, the German Israeli Foundation (Grant No. I-118-303.1-2018), the National 
Science Foundation through award DMR-2037654 and the US-Israel Binational Science Foundation
(BSF), and the Helmholtz International Fellow Award. U.K. was supported by the Raymond and
Beverly Sackler Faculty of Exact Sciences at Tel Aviv University and by the Raymond and Beverly Sackler
Center for Computational Molecular and Material Science.

\end{document}